\RequirePackage[2020-02-02]{latexrelease}
\documentclass[aps,prl,twocolumn,showpacs]{revtex4}
\usepackage{amssymb}
\usepackage{amsmath}
\usepackage{graphicx}
\usepackage{color}

\begin{document}
\title{Investigating the fast spectral diffusion of a quantum emitter in hBN using resonant excitation and photon correlations}
\author{Clarisse Fournier$^{1}$, Kenji Watanabe$^2$, Takashi Taniguchi$^3$, Julien Barjon$^{1}$, St\'ephanie Buil$^{1}$, Jean-Pierre Hermier$^1$, Aymeric Delteil$^1$}

\affiliation{$^1$ Universit\'e Paris-Saclay, UVSQ, CNRS,  GEMaC, 78000, Versailles, France. \\
$^2$ Research Center for Functional Materials, 
National Institute for Materials Science, 1-1 Namiki, Tsukuba 305-0044, Japan \\
$^3$ International Center for Materials Nanoarchitectonics, 
National Institute for Materials Science, 1-1 Namiki, Tsukuba 305-0044, Japan \\ {\color{white}--------------------} aymeric.delteil@usvq.fr{\color{white}--------------------} }

\date{\today }

\begin{abstract}
The ability to identify and characterize homogeneous and inhomogeneous dephasing processes is crucial in solid-state quantum optics. In particular, spectral diffusion leading to line broadening is difficult to evidence when the associated timescale is shorter than the inverse of the photon detection rate. Here, we show that a combination of resonant laser excitation and second-order photon correlations allows to access such fast dynamics. The resonant laser drive converts spectral diffusion into intensity fluctuations, leaving a signature in the second-order correlation function $g^{(2)}(\tau)$ of the scattered light that can be characterized using two-photon coincidences -- which simultaneously provides the homogeneous dephasing time. We experimentally implement this method to investigate the fast spectral diffusion of a color center generated by an electron beam in the two-dimensional material hexagonal boron nitride. The $g^{(2)}(\tau)$ function of the quantum emitter measured over more than ten orders of magnitude of delay times, at various laser powers, establishes that the color center experiences spectral diffusion at a characteristic timescale of a few tens of microseconds, while emitting Fourier-limited single photons ($T_2/2T_1 \sim 1$) between spectral jumps.
\end{abstract}

 \maketitle
\section{I. Introduction}

Dephasing processes are ubiquitous in the solid state, often limiting the practical applications of integrated quantum systems. Their characterization, although essential, is often challenging, owing to the great diversity of underlying physical mechanisms and therefore of the amplitude and timescales at which they can affect the system of interest. In the case of single-photon emitters (SPEs), such as color centers in wide gap crystals or self-assembled quantum dots, well-known examples of environment-induced effects acting on the spectral purity include fast dephasing due to phonons~\cite{Cassabois08, Flagg09, Tighineanu18, White21} and spectral diffusion caused by electrostatic fluctuations in the environment~\cite{Coolen09, Wolters13, Matthiesen14, Chen16, Spokoyny20}. The latter effect is typically slower than the emitter lifetime, thus leading to inhomogeneous broadening of the lineshape, and leaving a possibility to expose it by performing spectroscopy faster than its characteristic timescale~\cite{Kuhlmann13}. However, this method sets highly stringent conditions on the photon detection rate and is limited to slow phenomena. Roundabout methods have been demonstrated, based on photon interferometry~\cite{Brokmann06} or sub-linewidth filtering~\cite{Sallen10, Abbarchi12} -- but these techniques are in practice limited in terms of either temporal or spectral resolution.

Resonant laser excitation is a powerful way to perform spectroscopy at both high spectral resolution (limited by the laser linewidth) and time resolution (limited by the resolution of the photon counters). For a purely homogeneously broadened two-level system, the photon scattering rate $C$ is constant and only depends on the laser power and detuning. However, if the emitter experiences spectral fluctuations, $C$ will inherit these variations as a result of the time-dependent laser-emitter detuning. The induced intensity fluctuations will in turn leave a characteristic signature in the second-order correlation function $g^{(2)}(\tau)$, with a short-time bunching that depends on the laser power and detuning relative to the center of the inhomogeneous lineshape. In addition, $g^{(2)}(\tau)$ also provides other essential parameters such as the homogeneous coherence time $T_2$ -- that can be extracted from the near-zero-delay behavior. 
By including the influence of spectral diffusion, we show that the Rabi oscillations observed in the $g^{(2)}(\tau)$ function experience an additional effective damping, which is approximately independent of the spectral diffusion amplitude, but increases with the laser power. Only by accounting for this additional effect can the correct value of $T_2$ be extracted from the Rabi oscillations. Moreover, contrarily to interferometric techniques ~\cite{Brokmann06,Wolters13}, here detection of resonant photons is not needed and can be performed in phonon sidebands or replica, often available in solid-state quantum emitters.

We implement this approach to study a quantum emitter in the two-dimensional (2D) material hexagonal boron nitride (hBN). This wide-gap material is an attractive platform for quantum photonics in 2D materials due to the high quality of the quantum emitters that it can host~\cite{tran16, bourrelier16, martinez16}. Recently, a class of blue-emitting color centers (termed B-centers) that can be controllably generated~\cite{Fournier21,Gale22,shevitski19, Roux22} have been shown to display bright and stable optical transitions. Their low-temperature linewidth is, however, sizably broader than what is expected from independent measures of $T_2$, which has been found of the order of the excited state lifetime $T_1$~\cite{Horder22, Fournier22}. By measuring photon correlations at various powers and detunings, we here demonstrate that the excess linewidth can be attributed to fast spectral diffusion occuring at a rate of a few tens of kHz. Our results shed new light on the broadening mechanisms of this class of quantum emitters, and our analysis can be applied to other quantum optical systems.

\section{II. Methods}
\label{methods}

\subsection{A. Experimental setup and photoluminescence}

\begin{figure}[h]
  \centering
  \includegraphics[width=2.9in]{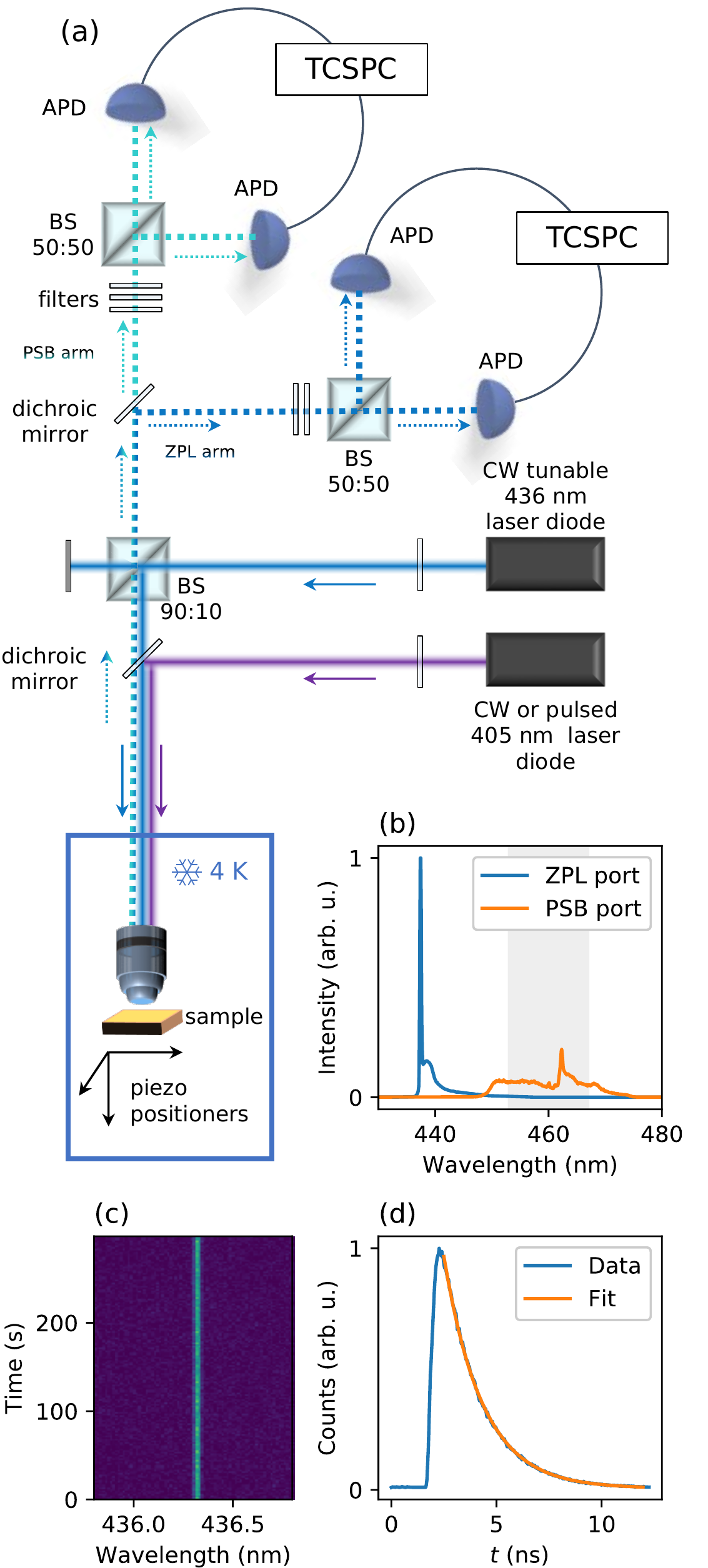}\\
  \caption{(a) Experimental setup. (b) PL spectra of SPEs through both output ports. The shaded area denotes the fluorescence filter in the PSB port. (c) PL spectrum as a function of time. (d) PL decay, yielding the emitter lifetime $T_1 = 1.83$~ns.}\label{fig1}
\end{figure}

Our sample consists in a single crystal of high-pressure, high-temperature grown hBN~\cite{Taniguchi07} exfoliated on a SiO$_2$(285~nm)/Si substrate and subsequently irradiated using an electron beam under 15~kV acceleration voltage~\cite{Fournier21}. The experimental setup is depicted in figure~\ref{fig1}(a): the sample is placed on piezoelectric positioners in a 4~K helium closed-cycle cryostat. A confocal microscope based on a 0.8 numerical aperture objective allows to address an individual B-center. The SPE is optically pumped using either a non-resonant [405~nm, pulsed or continuous-wave (cw)] or a resonant ($\sim$~436~nm, cw) laser. The resonant cw laser is frequency-locked to an external cavity and can be continuously tuned over the relevant wavelength range. Figure~\ref{fig1}(b) shows a low-temperature photoluminescence (PL) spectrum of the SPE. The emission features a zero-phonon line (ZPL) and a well-separated ($\sim$~160~meV) phonon sideband (PSB). The output port of the confocal microscope includes a dichroic mirror that separates the PSB from the ZPL. Both components are coupled to single-mode fibers and detected by avalanche photodiodes (APDs), with the PSB port being the detection channel when resonant excitation is used. The time stability of the ZPL can be observed Figure~\ref{fig1}(c), where no spectral fluctuations exceeding the spectrometer resolution of $\sim 100~\mu$eV can be detected. Figure~\ref{fig1}(d) shows the fluorescence decay of the emitter under pulsed excitation. An exponential fit provides the lifetime $T_1 = 1.83$~ns of the excited state.

\subsection{B. Resonant excitation of the emitter}

\begin{figure}[t]
  \centering
  \includegraphics[width=3.5in]{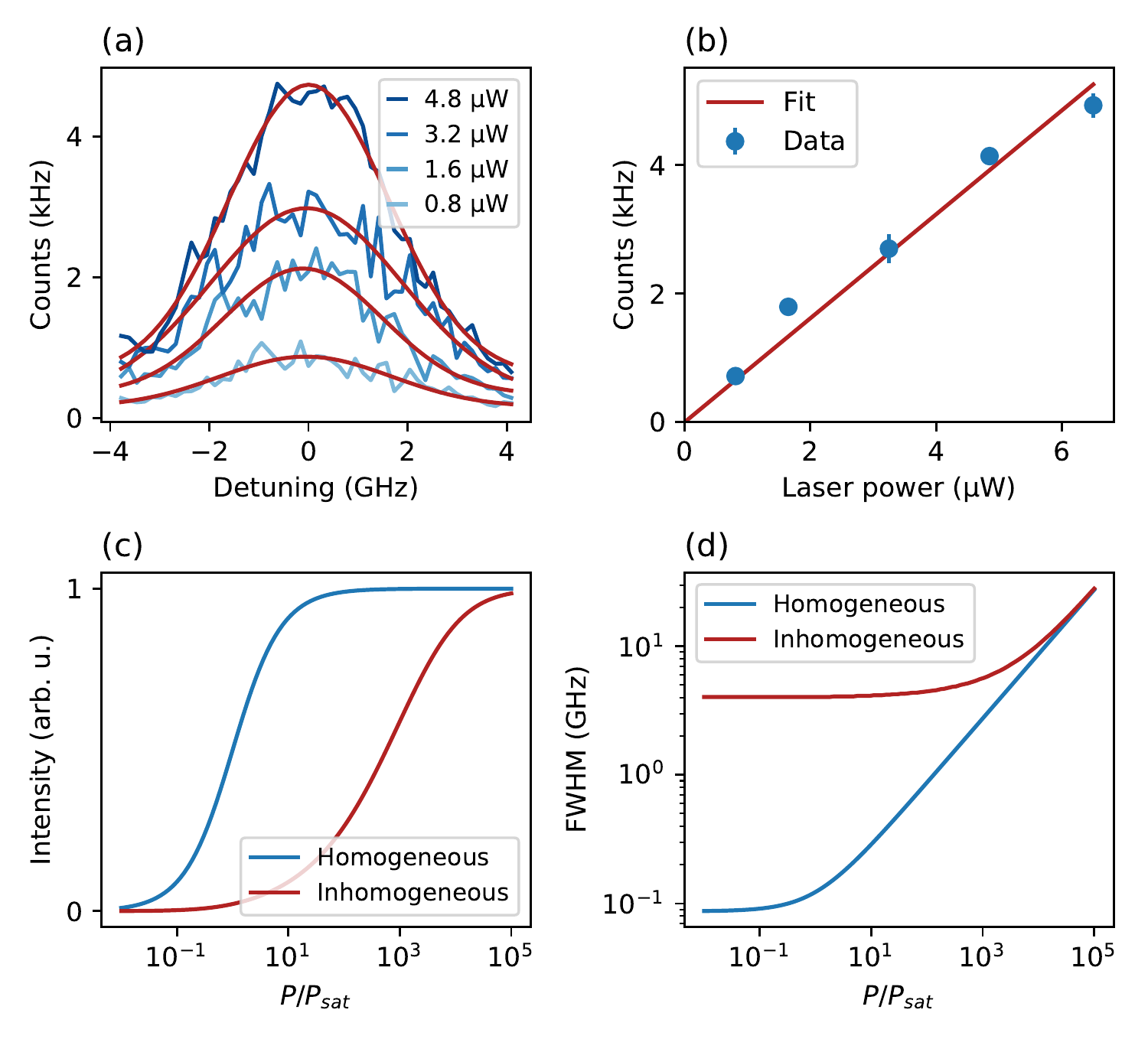}\\
  \caption{(a) Resonant laser scans of the SPE. The solid line are Gaussian fits to the data. (b) Maximum count rate as a function of the laser power. (c) Count rate as a function of the saturation parameter $P/P_\mathrm{sat}$, calculated in the homogeneous (blue line) and inhomogeneous (red line) case. (d) Linewidth of resonant laser scans calculated as a function of $P/P_\mathrm{sat}$ in the homogeneous (red line) and inhomogeneous (orange line) case.}\label{fig2}
\end{figure}

The response of the quantum emitter to a resonant laser is shown figure~\ref{fig2}(a) for various powers. Gaussian fits to the data provide the inhomogeneous linewidth of $4.0 \pm 0.3$~GHz, well above the natural linewidth (90~MHz) corresponding to the measured $T_1$~[fig.~\ref{fig1}(d)]. The maximum count rate as a function of the power is shown figure~\ref{fig2}(b).

We model the B-center as a two-level system with a transition energy $\hbar \omega_{ge}$ fluctuating in time, and following a Gaussian probability density $\mathcal{G}(\omega_{ge})$ of width $\Delta \nu = 4.0$~GHz. The homogeneous response is proportional to the excited state population~\cite{Loudon}:

\begin{equation}
C(\Omega_R,\omega_L,\omega_{ge}) \propto \rho_{ee} =  \dfrac{1}{2} \dfrac{\Omega_R^2  T_1/T_2}{(\omega_{ge}-\omega_L)^2 + T_2^{-2} + \Omega_R^2 T_1/T_2}
 \label{spectrum}
\end{equation}

where $\Omega_R$ is the Rabi frequency and $\hbar \omega_L$ the laser energy. This expression allows to define some important parameters, such as the saturation power -- the power at which the population reaches half its maximum at zero detuning --, which corresponds to $\Omega_R = 1/\sqrt{T_1T_2}$. The blue curve on figure~\ref{fig2}(c) displays the count rate on resonance and in the absence of spectral diffusion, as calculated from Eq.~\ref{spectrum}. It exhibits the well-known saturation of the two-level system at high power. The blue curve on figure~\ref{fig2}(d) displays the full width at half maximum (FWHM) of the Lorentzian response described by Eq.~\ref{spectrum}, which is given by $2\pi\Delta\nu_\mathrm{hom} = 2\sqrt{1 + \Omega_R^2T_1T_2}/T_2$. The increase of the FWHM when $P \gtrsim P_\mathrm{sat}$ is known as power broadening. In the presence of spectral diffusion, the expected photon emission rate as a function of the laser energy and power can then be numerically computed as the expectation value $\bar{C}(\Omega_R,\omega_L) = \int d \omega_{ge} \mathcal{G}(\omega_{ge})C(\Omega_R,\omega_L,\omega_{ge})$. The red curve on figure~\ref{fig2}(c) [resp.~\ref{fig2}(d)] displays the calculated count rate (resp. the linewidth) as a function of the saturation parameter $S = P/P_\mathrm{sat} = \Omega_R^2 T_1T_2$, for $T_2 = 2 T_1$. As long as the power-broadened linewidth $\Delta\nu_\mathrm{hom}$ stays below $\Delta \nu$, the lineshape does not vary much --even above saturation--, and the count rate stays approximately linear. Apparent saturation then occurs when the power broadening exceeds the inhomogeneous broadening. This can lead to a potentially large overestimation of the saturation power in such cases, if inferred as the power at which the count rate reaches half its maximum value. In our case, we find $P_\mathrm{sat} \approx 50$~nW from $T_1$, $T_2$ and $\Omega_R$ extracted from the experimental data in the following section.



\section{III. Second-order photon correlations}
\label{g2}

\subsection{A. Short-time correlations}

In the absence of spectral diffusion, a two-level atom driven by a cw laser exhibits antibunching and, above saturation, undergoes Rabi oscillations, which leave a characteristic signature in the second-order correlation function. At perfect resonance ($\omega_L = \omega_{eg}$), it is described by the following expression~\cite{Loudon}, accounting for both coherent effects and fast dephasing:

\begin{equation}
g^{(2)}(\tau) = 1 - e^{-\frac{T_1 + T_2}{2T_1 T_2} \tau} \left(   
\cos \Omega_R\tau + \frac{T_1 + T_2}{2 T_1 T_2 \Omega_R} \sin  \Omega_R\tau
\right)
 \label{shorttimecorrelations}
\end{equation}

This equation is widely used to fit experimental data of coherently driven solid-state quantum emitters~\cite{Flagg09, Horder22}, even in the presence of strong spectral diffusion. In such situation however, the laser-emitter detuning spans a range much broader than the emitter homogeneous linewidth, so that eq.~\ref{shorttimecorrelations} is not valid anymore. In the following, we generalize eq.~\ref{shorttimecorrelations} to include the effect of spectral diffusion.

At a fixed and finite detuning, eq.~\ref{shorttimecorrelations} is modified by replacing $\Omega_R$ by $\sqrt{\Omega_R ^2 + \delta^2}$, where $\delta = \omega_L - \omega_{ge}$ is the laser-emitter detuning~\cite{Rezai19}, yielding a more general expression of $g^{(2)}(\tau, \delta)$. Note that the two-fold coincidence rate also depends on $\delta$ since it varies quadratically with the count rate $C(\Omega_R, \delta)$ (as defined in eq.~\ref{spectrum}).

We now consider the presence of spectral diffusion, with an amplitude much larger than the homogeneous linewidth. If all values of the detuning $\delta$ are equally likely, each participates to the time-averaged coincidence rate with a relative weight that goes with the squared photon emission rate at this detuning, since $g^{(2)}$ is measured via two-photon coincidences. Therefore, the experimentally measured, time-averaged, $\tilde{g}^{(2)}(\tau) = \langle g^{(2)}(\tau, \delta) \rangle_\delta$ can be calculated as the weighted average of $g^{(2)}(\tau, \delta)$ over all possible detunings, namely:

\begin{equation}
\tilde{g}^{(2)}(\tau) \propto \int d\delta \ C(\Omega_R,\delta)^2 g^{(2)}(\tau, \delta) 
 \label{shorttimecorrelations_convolution}
\end{equation}

This result is independent of the shape and width of the inhomogeneous distribution as long as it exceeds the homogeneous linewidth, \textit{i.e.} $\Delta\nu \gg 2\sqrt{1 + \Omega_R^2T_1T_2}/T_2$, which is always the case in the present work. The influence of spectral diffusion on the behavior of $\tilde{g}^{(2)}(\tau)$ is investigated in detail in the Supplementary Information~\cite{Suppl}. It is found not to drastically modify the qualitative behavior of the oscillations, owing to the dependence in $C(\Omega_R,\delta)^2$ that favors contributions from close-to-zero-detuning components of the integrand. Nonetheless, integration of a continuum of non-resonant components leads to an additional damping of the Rabi oscillations. Therefore, using the zero-detuning eq.~\ref{shorttimecorrelations} to fit experimental data in a spectrally diffusive situation causes a sizable underestimation of $T_2$ --especially at high power-- and a (slight) overestimation of $\Omega_R$ by up to 20~\%. This bias is analyzed in detail in section~3 of the Supplementary information. It could explain power-dependent $T_2$ observed in prior work for quantum emitters experiencing spectral diffusion~\cite{Horder22, Martinez22}. In the intermediate case where $\Delta \nu \sim 2\sqrt{1 + \Omega_R^2T_1T_2}/T_2$, the exact inhomogeneous distribution has to be accounted for (see section~2 of the Supplementary Information~\cite{Suppl}).

\begin{figure}[t]
  \centering
  \includegraphics[width=3.5in]{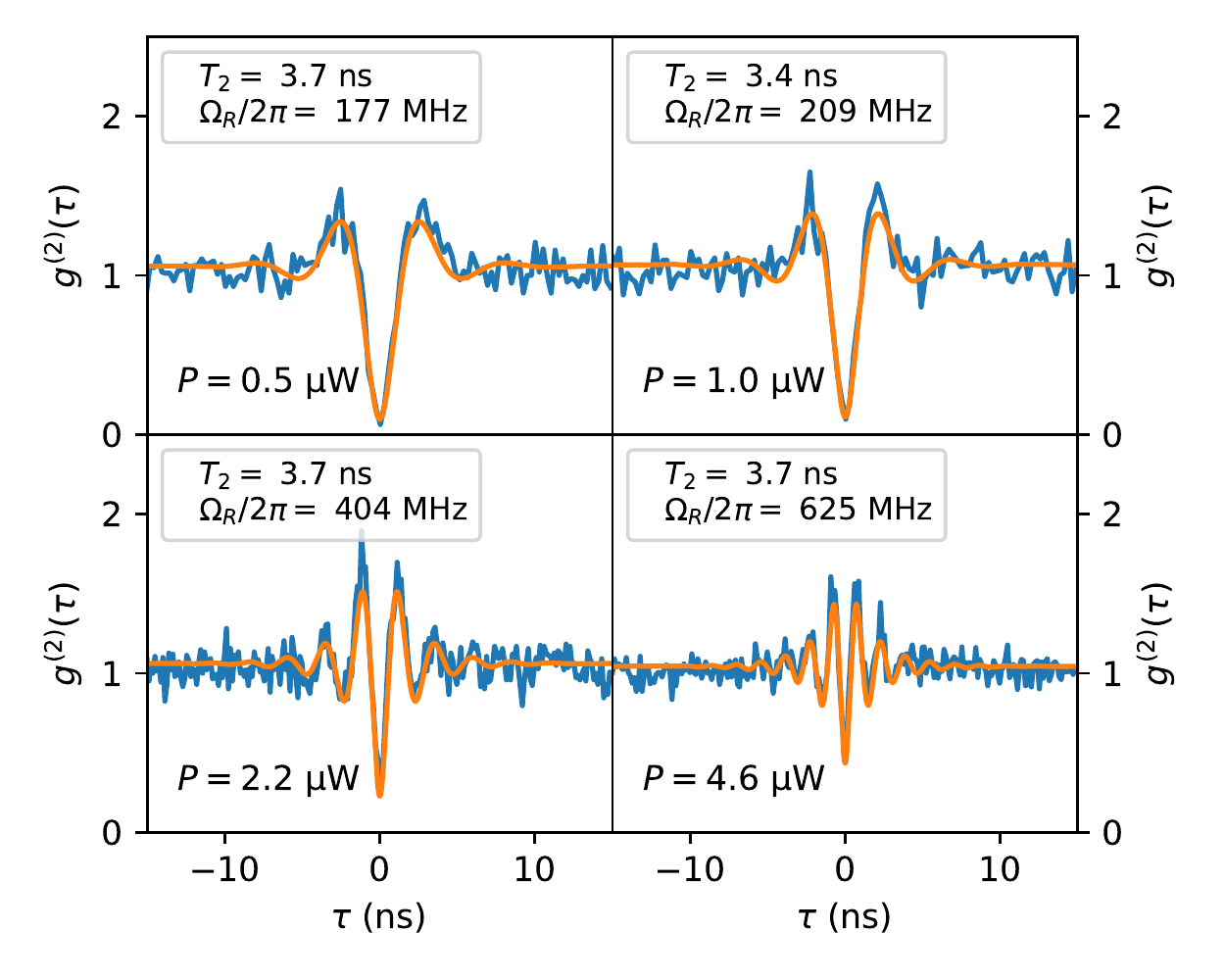}\\
  \caption{Blue lines: Short-time correlations at four different laser powers, revealing antibunching and damped Rabi oscillations. Orange lines: Fit to the data using eq.~\ref{shorttimecorrelations_convolution} that accounts for spectral diffusion.}\label{fig3}
\end{figure}

In order to measure the second-order correlation function of the emitter, the PSB output is detected in a Hanbury Brown and Twiss (HBT) setup. Figure~\ref{fig3} shows $g^{(2)}(\tau)$ as measured for $|\tau| < 15$~ns for four different laser powers. At these short delay times, the intensity correlation exhibits the expected behavior, with antibunching as well as damped oscillations. We fit the data with equation~\ref{shorttimecorrelations_convolution}, yielding $T_2 \sim 2 T_1$ at all powers. This indicates that the SPE emits transform-limited photons between spectral jumps. Transform-limited linewidth have been already observed in other families of hBN defects~\cite{Dietrich18, Akbari22}, although most studies of yellow to infrared color centers concluded to much shorter coherence times~\cite{Sontheimer17, Konthasinghe19,Spokoyny20}. We emphasize that the use of eq.~\ref{shorttimecorrelations} insted of eq.~\ref{shorttimecorrelations_convolution} in the fitting procedure yields slightly shorter values of $T_2$, ranging from 3.4~ns to 1.8~ns depending on the power. This is further discussed in section~4 of the Supplementary Information~\cite{Suppl}. Overall, these results provide a good example of a situation where homogeneous properties can be observed in a time-averaged measurement despite a much broader inhomogeneous broadening at large timescales. Other cases where homogeneous properties can be inferred in spite of inhomogeneously broadened energies include spectral hole burning~\cite{Palinginis03}, dynamical decoupling~\cite{Abella66, Press10} and mesoscopic ensembles of emitters~\cite{Delteil22}.

\subsection{B. Long-time correlations}

\begin{figure}[t!]
  \centering
  \includegraphics[width=3.18in]{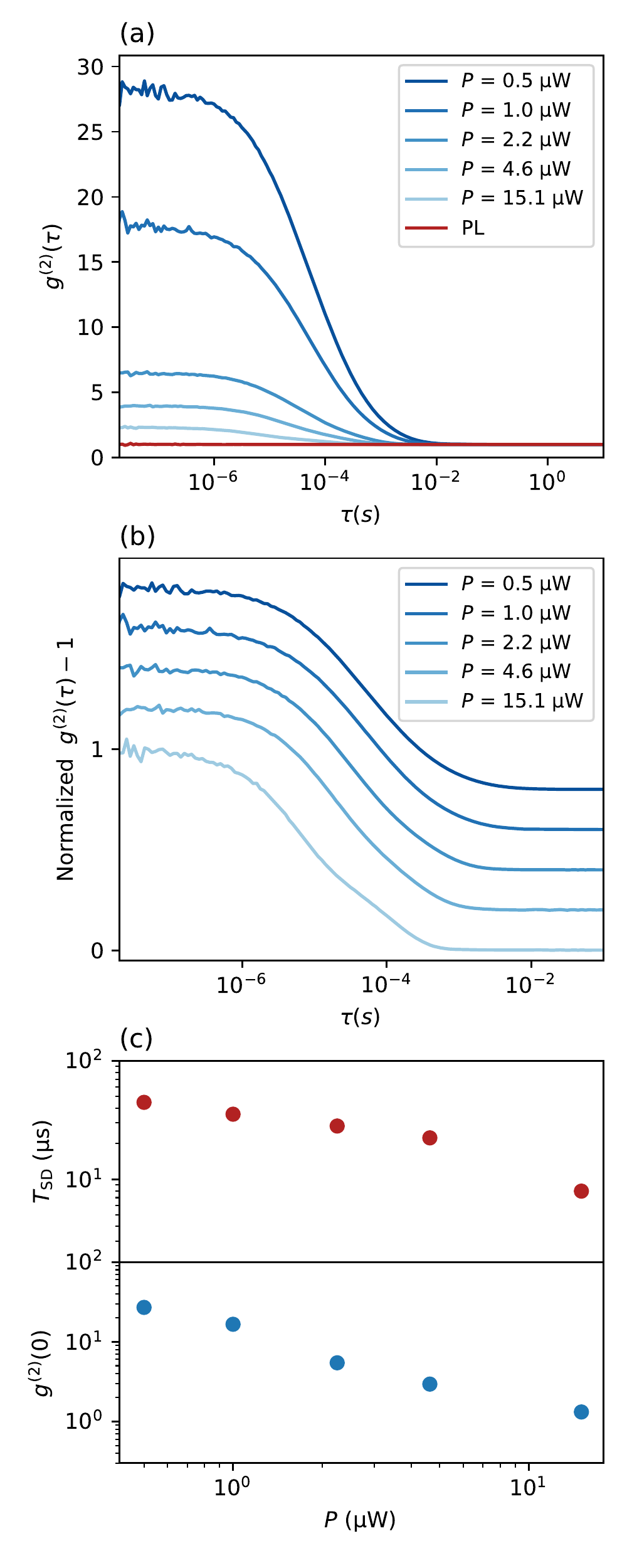}\\
  \caption{(a) Long-time $g^{(2)}(\tau)$ under non-resonant excitation (red line) and under resonant excitation at five different laser powers (blue lines). (b) Same data, but normalized and vertically shifted for clarity. (c) $T_\mathrm{SD}$ (upper panel) and $g^{(2)}_\mathrm{st}$ (lower panel) as a function of the laser power.}\label{fig4}
\end{figure}

The longer-time coincidences are processed in logarithmically scaled time bins~\cite{Laurence06} integrating delay times $\tau$ up to 10~s. The resulting coincidence rate is normalized by its long-time asymptote. The results for various laser powers, at zero detuning, are shown fig~\ref{fig4}(a) (solid lines). Photon bunching can be observed at short delays at all powers -- albeit with different magnitudes. The amount of bunching decreases with the delay time, reaching a plateau for $\tau \sim$~10~ms. No other variations in $g^{(2)}(\tau)$ are visible up to $\tau =$~10~s. To confirm that the statistics we observe originates from spectral diffusion and not from emitter blinking, we also measure $g^{(2)}(\tau)$ under non-resonant excitation [red curve in figure~\ref{fig4}(a)], which feature no bunching behavior at all. Incidentally, this confirms the validity of the two-level system approximation for this defect, in contrast to other hBN color centers~\cite{martinez16, Sontheimer17, Konthasinghe19}. The presence of blinking would yield bunching under both resonant and non-resonant excitation. On the contrary, spectral fluctuations do not turn into intensity fluctuations when the emitter is non-resonantly excited -- hence exhibits a constant $g^{(2)}(\tau)$.

To gain more insight into the dependence of $g^{(2)}(\tau)$ with the delay time and the laser power, on figure~\ref{fig4}(b) we plot the degree of bunching $g^{(2)}(\tau) -1$ as normalized by its maximum value. We can observe that their time dependence has a similar shape at all powers. However, the inflection point shifts to short times at high power. We define the spectral diffusion timescale $T_\mathrm{SD}$ as the time at which the degree of bunching decreases by half its maximum value, \textit{i.e.} $g^{(2)}(T_\mathrm{SD}) - 1 = (g^{(2)}_\mathrm{st} - 1)/2$, where we define $g^{(2)}_\mathrm{st}$ as the short-time bunching plateau of $g^{(2)}(\tau)$. The upper panel of figure~\ref{fig4}(c) shows $T_\mathrm{SD}$ as a function of the laser power. $T_\mathrm{SD}$ lies in the 10s of $\mu$s range, decreasing with the power from about 50~$\mu$s down to about 10~$\mu$s. This timescale is similar to the correlation time measured for other families of hBN quantum emitters using interferometric techniques~\cite{Sontheimer17, Spokoyny20}. The power dependence of $T_\mathrm{SD}$ indicates that the laser has an influence on the spectral diffusion. Moreover, the highest power curve suggests what appears like a splitting into two spectral diffusion processes. Finally, the lower panel of figure~\ref{fig4}(c) shows the short-time asymptote $g^{(2)}_\mathrm{st}$ as a function of the laser power. The amount of bunching strongly decreases with the power [Figure~\ref{fig4}(c), lower panel]. This is due to power broadening decreasing the ratio between homogeneous and inhomogeneous linewidths, as we argue in the following.

To gain more insight into the amount of bunching measured at $\tau \ll T_{SD}$ and its dependence on the laser power, we calculate the second-order correlation expected in the case of an inhomogeneously broadened two-level system described by eq.~\ref{spectrum}. Neglecting the quantum effects occuring at $\tau \lesssim T_1$ -- which have been treated in the previous section --, the  short-time asymptote $g^{(2)}_\mathrm{st}$ can be identified with the classical approximation of $g^{(2)}(0)$:

\begin{equation}
g^{(2)}_\mathrm{st} = \dfrac{\langle I^2 \rangle}{\langle I \rangle^2} = \dfrac{\int d \omega_{ge} \mathcal{G}(\omega_{ge})C(\Omega_R,\omega_L,\omega_{ge})^2}{\left[\int d \omega_{ge} \mathcal{G}(\omega_{ge})C(\Omega_R,\omega_L,\omega_{ge})\right]^2}
 \label{longtimecorrelations}
\end{equation}

\begin{figure}[t]
  \centering
  \includegraphics[width=3in]{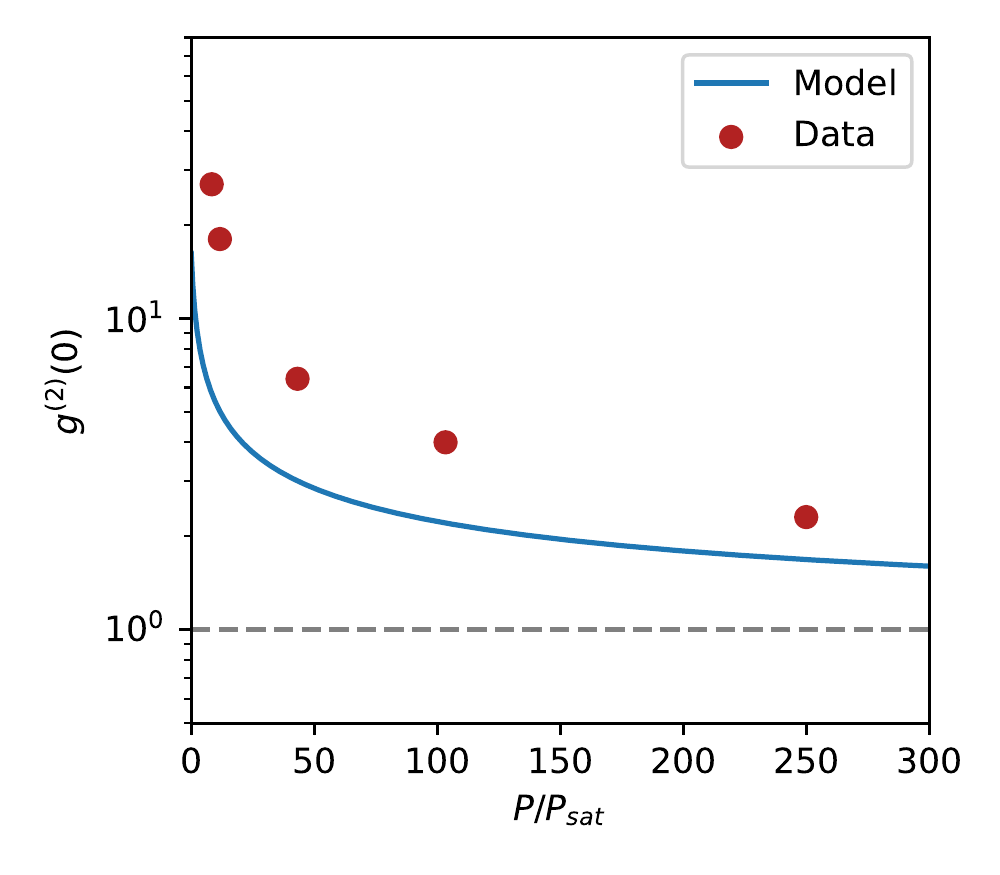}\\
  \caption{Solid line: $g^{(2)}_\mathrm{st}$ as a function of the laser power, as calculated using eq.~\ref{longtimecorrelations}. Red dots: Experimental values of $g^{(2)}_\mathrm{st}$ as a function of the laser power. Dashed line: High-power asymptote ($g^{(2)}_\mathrm{st} = 1$).}\label{fig5}
\end{figure}

Figure~\ref{fig5} shows the result calculated at zero detuning as a function of the laser power (blue curve), for $T_2 = 2 T_1$. At low power, the value of $g^{(2)}_\mathrm{st}$ is fixed by the ratio between the natural linewidth and the width of the inhomogeneous distribution. Intuitively, detection of a photon at time $t_1$ indicates that $\omega_{eg} \approx \omega_L$, which will stay so for an average duration $T_{SD}$. Therefore, the probability to detect a second photon at $t_2$ such that $t_2 - t_1 \ll T_{SD}$ is increased as compared to the random $\omega_{eg}$ case, which therefore generates bunching. At high power on the opposite, the power-broadened homogeneous linewidth exceeds $\Delta \nu$ such that the intensity fluctuations vanish -- which leads to $g^{(2)}_\mathrm{st}(P \rightarrow \infty) = 1$.

The experimental value of $g^{(2)}_\mathrm{st}$ is plotted on the same figure (red dots on figure~\ref{fig5}). The qualitative trend follows the model -- however, we observe more bunching than expected from the calculation. The origin of this excess bunching is not understood, and could involve a complex interplay between the transition energy and the laser detuning. Prior work on other semiconductor systems have unveiled various laser-induced spectral jumps mechanisms, which tend to bring the system off-resonant by means of environment modifications~\cite{Fernee12, Hogele12}. Such negative laser-emitter correlations are expected to increase photon bunching. We note that the long integration time ($> 24$~h at the lowest powers) could also have an influence on the spectral diffusion.

\section{IV. Conclusion}

We have shown that photon correlations under resonant excitation constitutes a powerful tool to investigate spectral fluctuations of solid-state quantum emitters, combining high temporal and spectral resolutions. The short-time correlations exhibit a quantum coherent behavior that allows to extract $T_2$ and $\Omega_R$. The effect of spectral diffusion is accounted for, and results in a faster damping of the Rabi oscillations. On the other hand, the long-time correlations provides information about the spectral diffusion dynamics. We have applied this technique to an individual B-center in hBN, showing that it can emit homogeneous ensembles of $T_\mathrm{SD}/T_1 \sim 10^4$ consecutive photons that are close to transform-limited ($T_2/2T_1 \sim 1$), before experiencing spectral jumps. This number could be further improved by both reducing the photon lifetime through Purcell effect and reducing the environment noise (with for instance gated devices~\cite{Akbari22}). 

The data generated in this study are available~\cite{data}.

\section{Acknowledgments}
The authors thank A.~Pierret and M.~Rosticher for flake exfoliation, and S.~Roux for sample irradiation. This work is supported by the French Agence Nationale de la Recherche (ANR) under reference ANR-21-CE47-0004-01 (E$-$SCAPE project). This work also received funding from the European Union’s Horizon 2020 research and innovation program under Grant No. 881603 (Graphene Flagship Core 3). K.W. and T.T. acknowledge support from JSPS KAKENHI (Grant Numbers 19H05790, 20H00354 and 21H05233).

\pagebreak
~
\newpage

\onecolumngrid
\begin{center}
  \textbf{\large Supplementary Material\\~\\Investigating the fast spectral diffusion of a quantum emitter in hBN using resonant excitation and photon correlations}\\[.2cm]
  Clarisse Fournier$^{1}$, Kenji Watanabe$^2$, Takashi Taniguchi$^3$, Julien Barjon$^{1}$, St\'ephanie Buil$^{1}$, Jean-Pierre Hermier$^1$, Aymeric Delteil$^1$\\[.1cm]
  {\itshape \small $^1$ Universit\'e Paris-Saclay, UVSQ, CNRS,  GEMaC, 78000, Versailles, France. \\
$^2$ Research Center for Functional Materials, 
National Institute for Materials Science, 1-1 Namiki, Tsukuba 305-0044, Japan \\
$^3$ International Center for Materials Nanoarchitectonics, 
National Institute for Materials Science, 1-1 Namiki, Tsukuba 305-0044, Japan \\
{\color{white}--------------------} aymeric.delteil@usvq.fr{\color{white}--------------------} \\}
(Dated: \today)\\[1cm]
\end{center}

\setcounter{equation}{0}
\setcounter{figure}{0}
\setcounter{table}{0}
\setcounter{page}{1}
\renewcommand{\theequation}{S\arabic{equation}}
\renewcommand{\thefigure}{S\arabic{figure}}
\renewcommand{\bibnumfmt}[1]{[S#1]}
\renewcommand{\citenumfont}[1]{S#1}

\section*{Short-time second-order correlation function}
\vspace{0.5cm}
\textbf{1. Without spectral diffusion \\}

\begin{figure*}[h!] 
\centering
\includegraphics[width=0.5\textwidth]{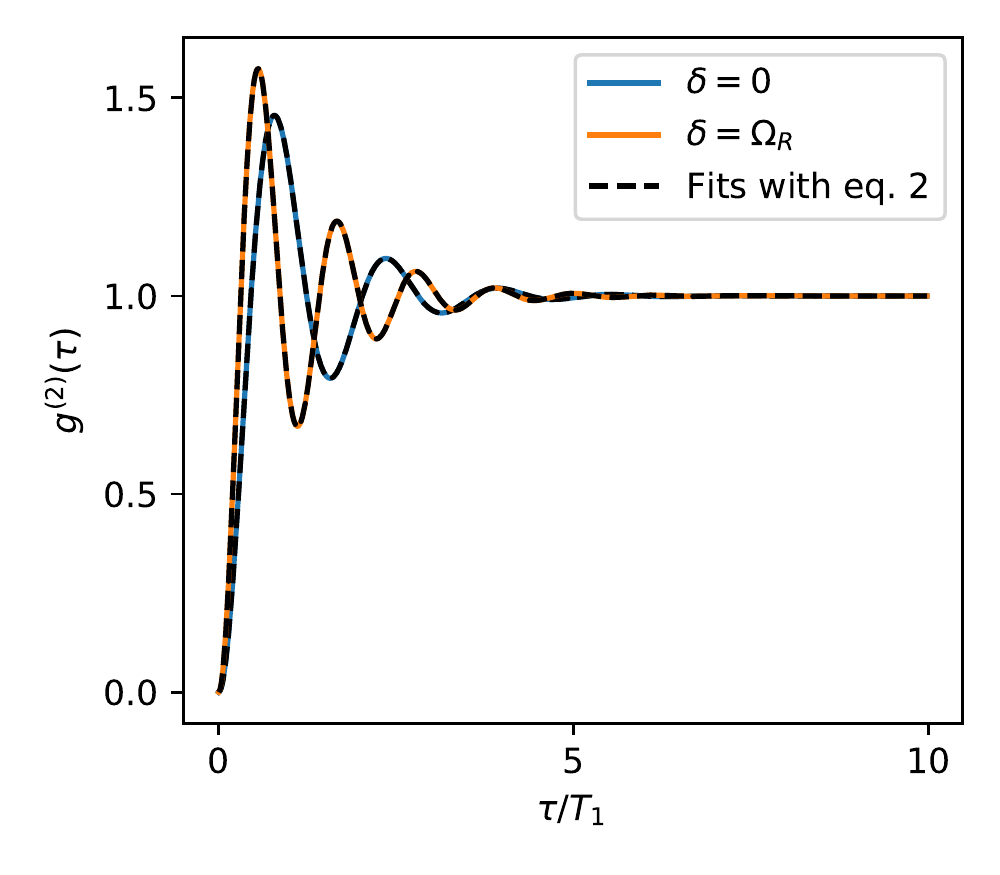}
\caption{Solid lines: $g^{(2)}(\tau)$ calculated based on numerical integration of the master equation for $\Omega_R = 4 \Gamma_1$ and $\Gamma_2 = \Gamma_1$ at two detunings $\delta = 0$ and $\delta = \Omega_R$. Dashed lines: Fit using eq.~2 of the main text.} \label{figS1}
\end{figure*}

In the absence of spectral diffusion, the second-order correlation function reads

\begin{equation}
g^{(2)}(\tau) = 1 - e^{-\frac{T_1 + T_2}{2T_1 T_2} \tau} \left(   
\cos \sqrt{\Omega_R^2 + \delta^2}\tau + \frac{T_1 + T_2}{2 T_1 T_2 \sqrt{\Omega_R^2 + \delta^2}} \sin  \sqrt{\Omega_R^2 + \delta^2}\tau
\right)
 \label{shorttimecorrelationsSupp}
\end{equation}

as established from optical Bloch equations and the quantum regression theorem~\cite{Loudon}. This equation simplifies to eq.~2 of the main text at zero atom-laser detuning -- which can also be used to fit the finite detuning behavior. It then provides $\tilde{\Omega}_R = 
\sqrt{\Omega_R^2 + \delta^2}$ instead of $\Omega_R$. We illustrate this by generating $g^{(2)}$ data using numerical integration of the master equation~\cite{Breuer, Gardiner} (plain curves on figure~\ref{figS1}) at two different detunings, and subsequently fitting them using main text eq.~2 (dashed curves on figure~\ref{figS1}).

\vspace{0.5cm}
\newpage
\textbf{2. Accounting for spectral diffusion}
\vspace{0.5cm}

The time-averaged effective $g^{(2)}(\tau)$ is calculated as the (continuous) sum over a range of detunings that sufficiently exceeds the linewidth, weighted by the squared count rate $C(\delta)$ as defined in eq.~1 of the main text. This leads to eq.~3 of the main text.

In a case where the inhomogeneous distribution $S(\hbar \omega)$ has a width that is comparable with the homogeneous linewidth $2/T_2 \sqrt{1+\Omega_R^2T_1T_2}$, a more general expression has to be used:

\begin{equation}
\tilde{g}^{(2)}(\tau) \propto \int d\omega_{eg} \ S(\hbar \omega_{eg}) C(\Omega_R,\omega_{eg} - \omega_L)^2 g^{(2)}(\tau, \omega_{eg} - \omega_L) 
 \label{shorttimecorrelations_convolutionSupp}
\end{equation}

with $\hbar \omega_L$ is the laser energy, and where the spectral shape $S(\hbar \omega)$ acts as a probability density distribution for the emitter energy $\hbar \omega_{eg}$.

In our range of experimental parameters, we always have $\Delta \nu \gg 2/T_2 \sqrt{1+\Omega_R^2T_1T_2}$. We have also verified that the more accurate description of eq.~\ref{shorttimecorrelations_convolutionSupp} produces no perceptible difference in this range of parameters. Therefore, in the following we use eq.~3 of the main text as our model for the spectrally-diffusive emitter.

\begin{figure*}[h!] 
\centering
\includegraphics[width=0.8\textwidth]{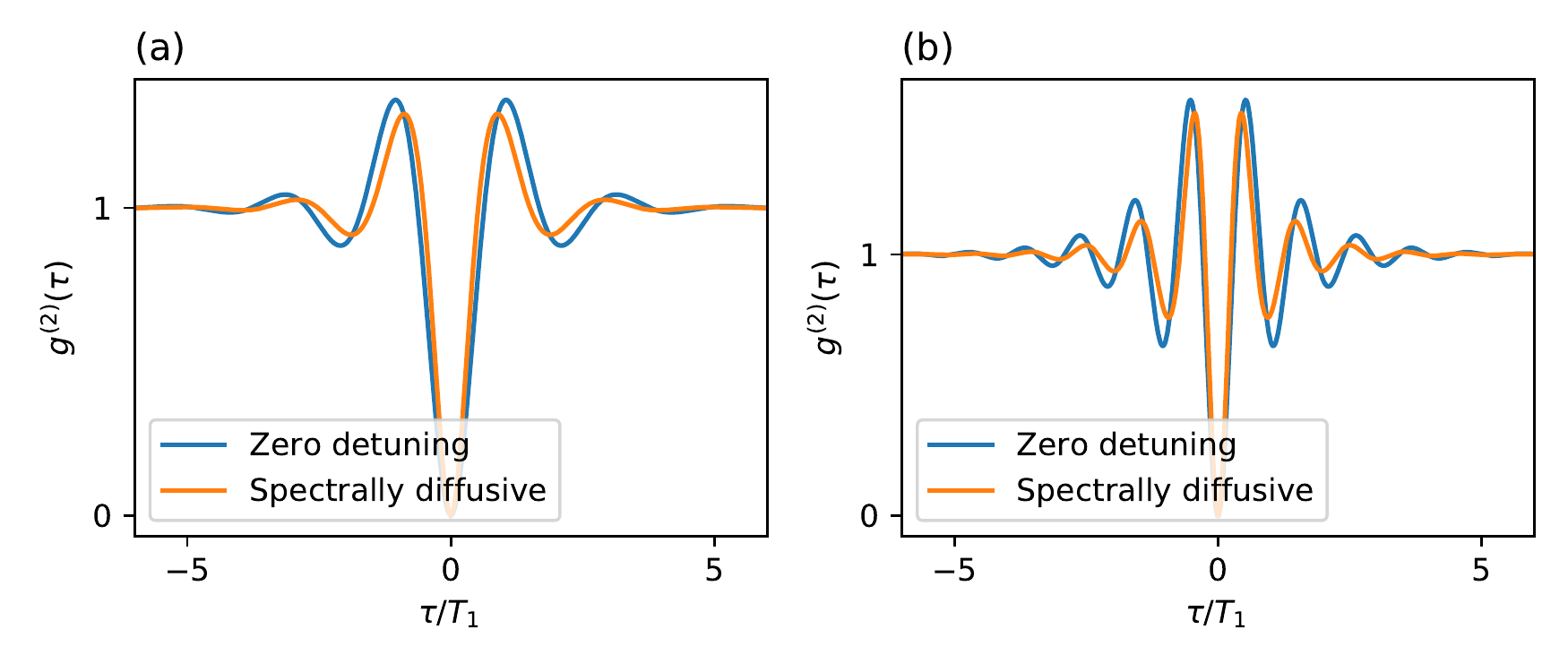}
\caption{(a) Zero-detuning (blue curve) and spectrally diffusive (orange curve) $g^{(2)}(\tau)$ with $\Omega_R = 3 \Gamma_1$ and $\Gamma_2 = \Gamma_1$. (b) Same as (a), with $\Omega_R = 6 \Gamma_1$.} \label{figS2}
\end{figure*}

Figure~\ref{figS2} shows $g^{(2)}(\tau)$ as calculated at two different powers, without and with spectral diffusion (i.e. based on eq.~2 and eq.~3 of the main text). It can be seen that the presence of spectral diffusion does not qualitatively modify the behavior of $g^{(2)}(\tau)$, which exhibits damped Rabi oscillations in both cases. However, the spectrally diffusive case shows a faster damping and a slightly shorter period of the oscillations.

The fact that the qualitative aspect of the short-time $g^{(2)}(\tau)$ is not drastically modified by spectral diffusion is remarkable. However, this observation can give the false impression that fitting with the zero-detuning expression (eq.~2 of the main text) is legitimate and directly provides the relevant physical quantities. Nonetheless, the quantitative modifications shown figure~\ref{figS2} suggest that the fitting procedure would provide biased values of $T_2$ and $\Omega_R$. 

\vspace{0.5cm}
\newpage
\textbf{3. Influence of the spectral diffusion on the fit parameters}
\vspace{0.5cm}

\begin{figure*}[h!] 
\centering
\includegraphics[width=0.8\textwidth]{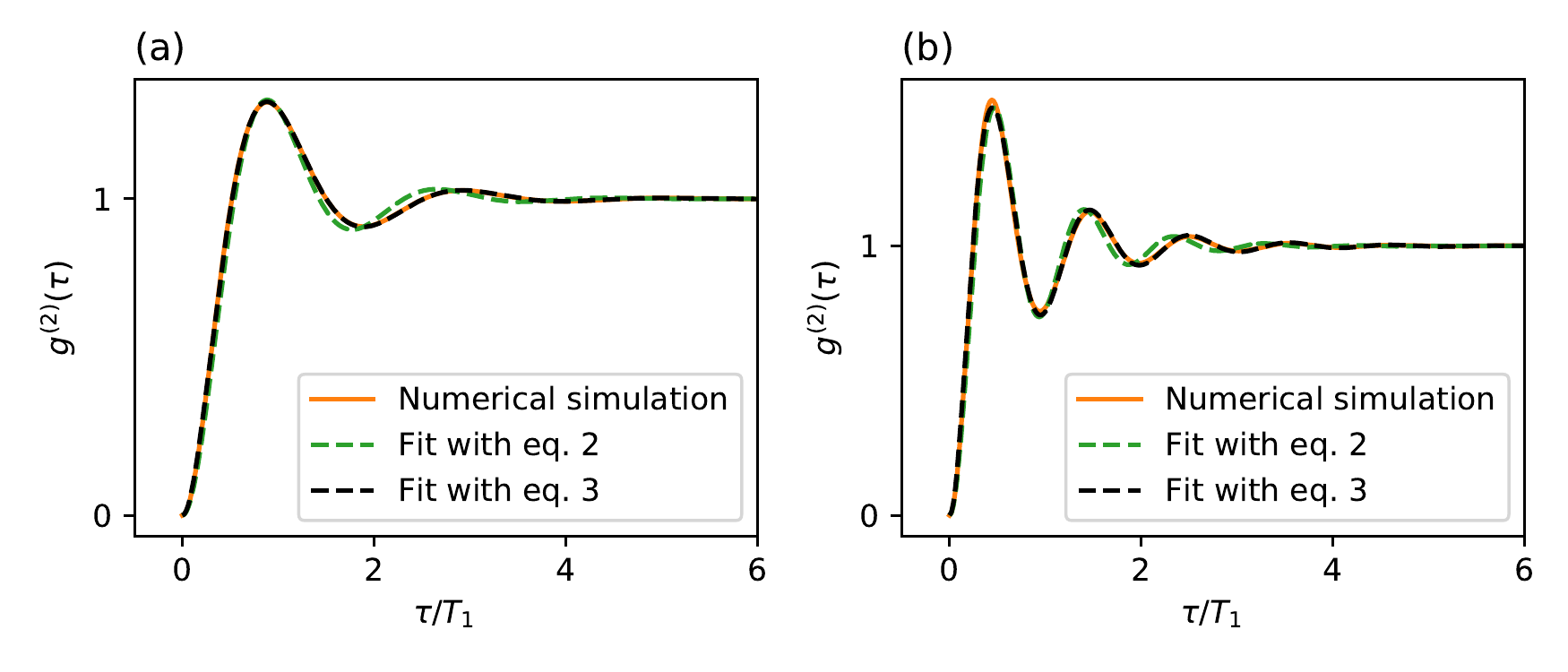}
\caption{(a) Spectrally diffusive $g^{(2)}(\tau)$ (orange curve) with $\Omega_R = 3 \Gamma_1$ and $\Gamma_2 = \Gamma_1$. Green dashed line: Fit with eq.~2. Black dashed line: Fit with eq.~3. (b) Same as (a), with $\Omega_R = 6 \Gamma_1$.} \label{figS3}
\end{figure*}

We illustrate this deviation by generating diffusive curves with the parameters of fig.~\ref{figS2} using numerical integration of the master equation, and subsequently fitting them using both models (\textit{i.e.} eq.~2 and eq.~3 of the main text). Figure~\ref{figS3} shows the fit result. Both zero-detuned and spectrally diffusive models fit the numerical simulation well. However, while the latter case provides the correct values for the physical quantities [\textit{i.e.} $T_2/T_1 = 1$ and $\Omega_R T_1 = 3$ (left panel) and 6 (right panel)] that were used as inputs in the numerical simulation, on the other hand the zero-detuning model yields $T_2/T_1 = 0.61$ and $\Omega_R T_1 = 3.55$ (left panel); and $T_2/T_1 = 0.54$ and $\Omega_R T_1 = 6.71$ (right panel) in place of the previously mentioned values.

\begin{figure*}[h!] 
\centering
\includegraphics[width=0.6\textwidth]{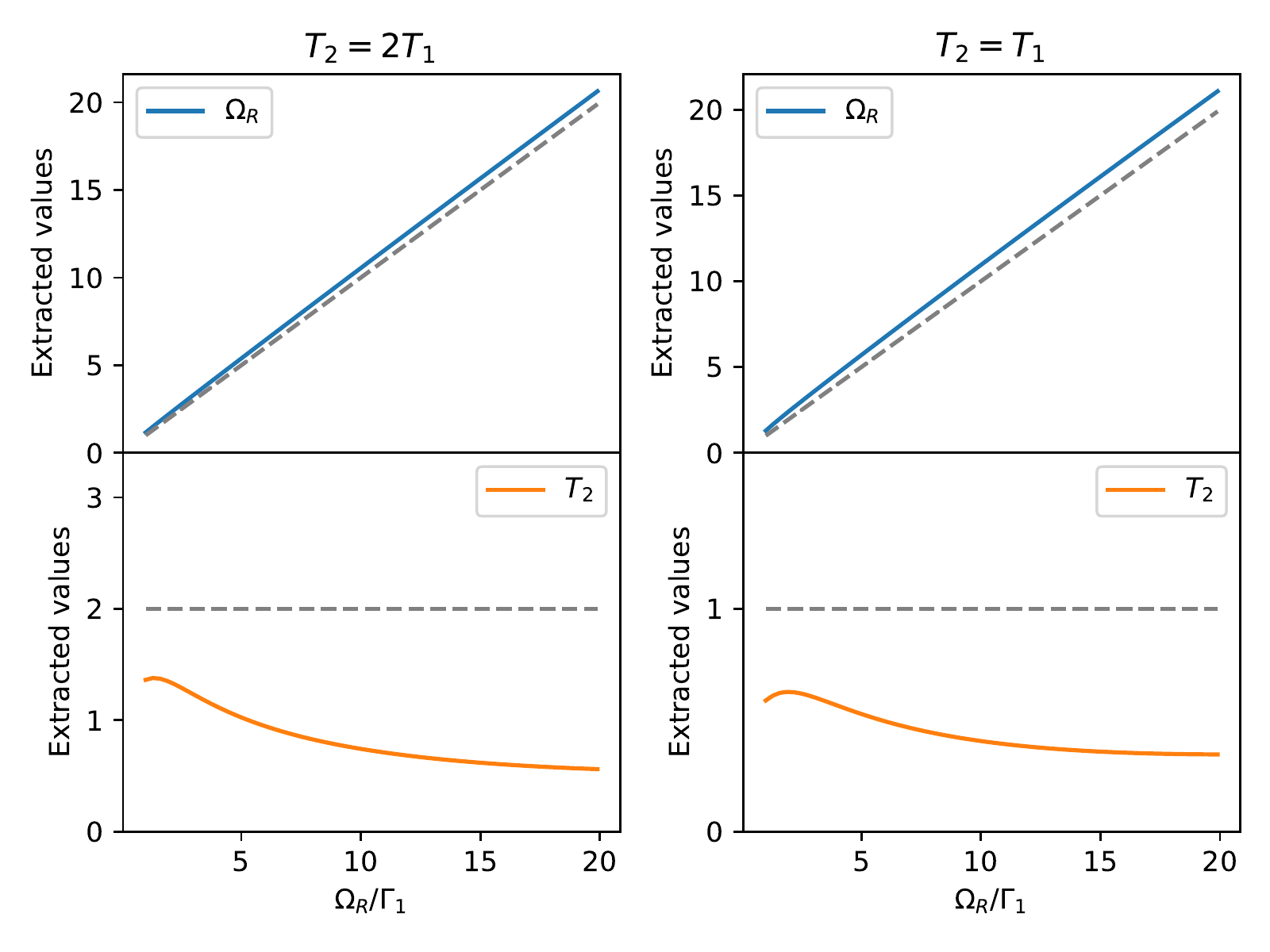}
\caption{Fitting results of $\Omega_R$ (blue curves) and $T_2$ (orange curve) using eq.~2, as a function of the laser Rabi frequency $\Omega_R$, in the case $T_2 = 2T_1$ (left panel) and $T_2 = T_1$ (right panel). The dashed gray lines indicate the input values.} \label{figS4}
\end{figure*}

In order to investigate further the bias originated from fitting Rabi oscillations from spectrally diffusive emitters with the zero-detuning eq.~2, we generate $g^{(2)}(\tau)$ using eq.~3 for a range of parameters, and subsequently fit them using eq.~2. Figure~\ref{figS4} shows the results for two different values of $T_2$ as a function of the Rabi frequency. These results show that the bias in the estimation of the Rabi frequency does not depend much on the laser power. On the contrary, the higher the power, the more $T_2$ is underestimated. This is related to the fact that, at higher power, power broadening increases the range of detuning actively participating to the overall shape, thereby increasing the effective damping of the oscillations. As mentioned in the main text, we propose that this mechanism could participate to the commonly observed power-dependent dephasing~\cite{Horder22, Martinez22}.

\vspace{0.5cm}
\newpage
\textbf{4. Experimental data}
\vspace{0.5cm}

\begin{figure*}[b!] 
\centering
\includegraphics[width=0.6\textwidth]{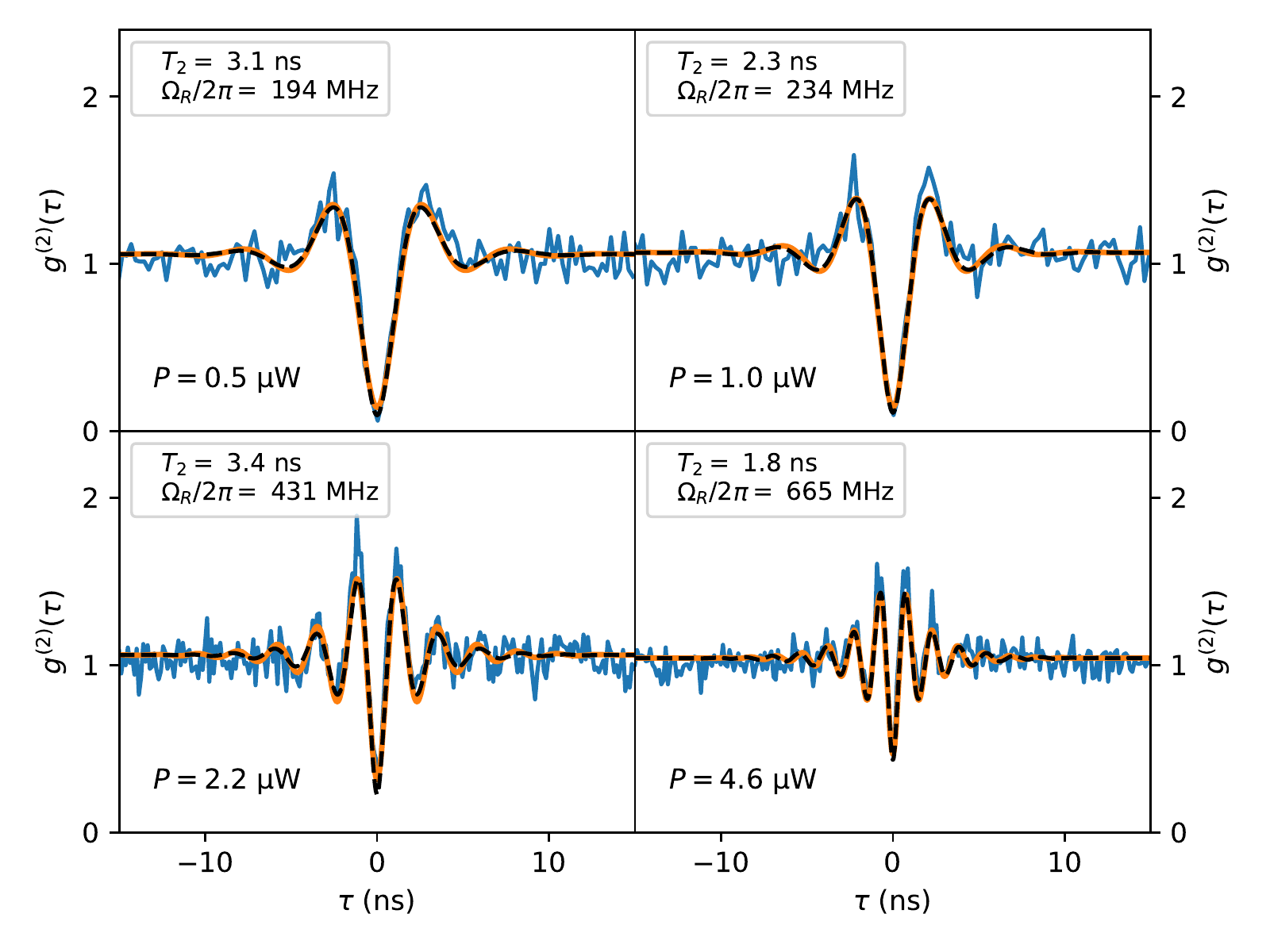}
\caption{Blue curves: Same data as in figure~3 of the main text. Orange curves: Fit results using main text eq.~2 instead of eq.~3. Boxes: Extracted values from the fit with eq.~2. Dashed lines: Fit results using main text eq.~3, as also plotted on figure~3 of the main text.} \label{figS5}
\end{figure*}

In this section, we fit the dataset of the main text figure~3 using main text eq.~2 -- \textit{i.e.} by neglecting the effect of spectral diffusion. Figure~\ref{figS5} displays the fit results in this case, together with the fits by main text eq.~3 (already shown on main text figure~3). It can be seen that both models fit equally well the data. However, as evidenced in the previous sections, neglecting spectral diffusion leads to an underestimation of the emitter coherence time, in particular at high laser power. On the opposite, when including the effect of spectral diffusion, our whole dataset can be well reproduced without the need to invoke fast dephasing processes (see fig.~5 of the main text).

\end{document}